\documentclass[12pt,preprint]{aastex}

\newcommand{\kms}{km\thinspace s$^{-1}$}

\begin{document}

\title{Jets and Tori in Proto-Planetary Nebulae}


\author{P. J. Huggins}
\affil{ Physics Department, New York University, 4 Washington Place,
 New York, NY 10012}
\email{patrick.huggins@nyu.edu}



\begin{abstract}

We investigate the time sequence for the appearance of jets and
molecular tori in the transition of stars from the Asymptotic Giant
Branch to the planetary nebula phase. Jets and tori are prominent
features of this evolution, but their origins are uncertain. Using
optical and millimeter line kinematics, we determine the ejection
history in a sample of well-observed cases.  We find that jets and
tori develop nearly simultaneously. We also find evidence that jets
typically appear slightly later than tori, with a lag time of a few
hundred years. These characteristics provide strong evidence that jets
and tori are physically related, and they set new constraints on
theories of jet formation.  The ejection of a discrete torus followed
by jets on a short time scale favors the class of models in which a
companion interacts with the central star. Models with long time
scales, or with jets followed by a torus, are ruled out.

\end{abstract}




\section{Introduction}

Planetary nebulae (PNe) are observed to exhibit remarkably complex
morphologies that are not well understood \citep[e.g.,][]{bal02,
mei04}.  Point symmetry is the most striking characteristic of the
nebulae and is known to be produced by the action of high velocity
jets which emanate from the central star system and interact with the
circumstellar gas. The jets are typically bipolar, but may show
multiple components or changes in direction. They are widely seen in
young PNe and proto-PNe \citep[e.g.,][]{sah98} and, in a few
cases, in the final stages of the Asymptotic Giant Branch (AGB)
\citep[e.g.,][]{ima02}.  The origin of the jets is, however, uncertain.

Four different types of scenario have been proposed for the formation
of the jets: a magnetic wind from single stars \citep{gar05}; an
accretion disk around a binary companion, fed by the mass-loss of the
primary \citep{mor87,sok00}; a magnetic (possibly explosive) wind from
the primary, spun up by a companion \citep[e.g.,][]{nor06,mat06}, and
an accretion disk around the core of the primary, formed by the
overflow or break-up of a binary companion during or after a common
envelope phase \citep{sok94,sok96,rey98,nor06}. In all these cases the
region of jet launching is too small to be resolved by current
observations. Our constraints on the process of jet formation are,
therefore, necessarily indirect, based on their large scale properties
and our understanding of the environment in which they form.

A second, extremely common characteristic of PNe and proto-PNe is a
region of equatorially enhanced mass loss. This often forms a
relatively well defined structure, variously described as a ring, or
torus.  The mass of this component may be large, $\ga 0.1$~M$_\odot$
with corresponding mass loss rates $\ga 10^{-4}$~M$_\odot$~yr$^{-1}$
\citep[e.g.,][]{hug04}, $\ga 1$--2 orders of magnitude larger than
typical mass loss rates on the upper AGB. The origin of this enhanced mass
loss is not known for certain, but it terminates the evolution of the
star on the AGB and occurs in the same transition phase as the first
appearance of prominent jets.

In this paper we explore possible relations between the ejection of
the tori and the development of jets. We determine the timing of their
appearance as accurately as possible using the kinematic structure of
a sample of well-observed transition objects, and use the results to
evaluate different jet formation scenarios.

\section{Observations}

In order to investigate the time sequence for the appearance of jets
and tori we consider transition objects whose dimensions, kinematics,
and orientation are sufficiently well characterized by observations
that we can estimate when these features formed. The sample is listed
in Table~1.  It includes late AGB stars, proto-PNe, and young PNe,
although the status of M~2-9, which is usually included among young
PNe is uncertain; it may be a symbiotic system, e.g.,
\cite{cor04b}. For convenience, we shall call the objects proto-PNe as
a group. Column (2) of the table gives the adopted distances for later
use.

The objects in Table~1 were selected on the basis of the following
criteria: they show well defined jets and tori; the kinematic
structure of the tori has been observed in millimeter CO emission at
high angular resolution; and the kinematics of the jets have been
observed in molecular lines, or at optical wavelengths. In fact, the
sample includes most of the transition objects that have so far been
mapped at high resolution in CO; we have omitted only a few cases in
which the jets or equatorial regions are complex, e.g., AFGL~2688
\citep{cox00}, or the geometry is poorly determined. The emphasis here
on molecular line observations of the tori is important for providing
robust reconstructions of the ejection history.  There are, of course,
many optical observations of the equatorial regions of fully formed
PNe, but in these cases the dynamical effects of pressure in the
ionized gas make reconstruction of the ejection history subject to
large systematic uncertainty.

Table 1 summarizes the data on the sample objects.  Columns (3)--(4)
give estimates for the radii and expansion velocities of the
equatorial tori ($r_t$ and $v_t$) for each object, and columns
(6)--(7) give the radial extents and velocities for the jets ($r_j$
and $v_j$).  The observations on which these values are based are
reported in the references in the table. The adopted values are taken
from spatio-kinematic models or data summaries by the authors cited,
or are based on new estimates made from maps given in the papers.  The
observations are corrected for projection using data on the
inclination angles (which we take to be the angle between the axis of
the torus or the jets, to the line of sight) given in the appendix.
The values of $r_t$ and $r_j$ are calculated using the distances given
in column (2) of Table~1.

The data for the tori are based on interferometric CO observations.
The values of $r_t$ refer to characteristic radii of the tori. In
cases where the inner hole is resolved by the observations, $r_t$ is
the radius of peak CO emission or the mean of inner and outer radii
from model fits by the authors. If the inner hole is marginally
resolved or not resolved, $r_t$ gives the radius modeled by the
authors, or the half-intensity radius estimated from the CO emission
maps.  The data for the jets are based on molecular line or optical
observations. The values of $r_j$ and $v_j$ refer to the prominent jet
heads farthest from the center of the system. These data are typically
averages for oppositely directed jets, but only one is used if the
data for it are superior.

The data on the radial extents and velocities of the jets and tori are
used to determine the expansion times given by $t_j=r_j/v_j$ and
$t_t=r_t/v_t$. These are listed in columns (5) and (8) of Table~1.
For three objects (He~3-1475, M~2-9, and KjPn~8), we have preferred to
use expansion times for the jets determined from optical proper
motions, because they are generally more accurate and are independent
of the inclination angle.  In these cases $t_j=\phi/\dot\phi$ where
$\phi$ is the angular distance of the jet head from the central star
and $\dot\phi$ is its time derivative.

In summary, the expansion times of the tori are all determined from
molecular line observations, while the expansions times of the
jets are from molecular line kinematics, optical line kinematics,
or optical proper motions. In two cases, M~1-16 and He~3-1475, there
are measurements of additional jet heads closer to the central star
system. Information on these and further details of the observations
are given in the appendix.

\section{Timing}

\subsection{Near Simultaneity}

We use the expansion times of the jets and tori as estimates for the
time since they were ejected from the central star systems. The large
masses and low expansion velocities of the dense molecular tori mean
that their expansion times ($t_t$) provide excellent estimates for
their travel times.  For the jets, there is a good deal of
observational evidence from detailed studies of individual objects for
approximately Hubble-type outflows that also suggest constant
velocity, ballistic motions \citep[e.g.,][]{alc01,cor04,uet06}. We
assume that this applies to the jet heads considered here. We discuss
this assumption further in \S3.2.

Fig.~1 compares the expansion times of the jets with the expansion
times of the tori. The squares denote cases where the torus is
resolved, and the diamonds denote cases where it is not well
resolved. The solid line shows the relation $t_j = t_t$.

One major uncertainty that affects the location of the data points in
Fig.~1 is caused by uncertain distances to the nebulae.  Apart from
the exceptions noted below, this affects both $t_j$ and $t_t$ in the
same way: it moves the plotted points parallel to the line $t_j=t_t$
in the figure and does not appreciably affect our discussion.

A second important effect that affects the location of the data points
in Fig.~1 is uncertainty in the inclination angles of the jets and
tori.  The adopted values and estimates of the uncertainties are given
in the appendix.  In most cases the values of $t_t$ and $t_j$ vary
with inclination as $\sin \theta$ and $\cot \theta$, respectively, and
the error bars shown in Fig.~1 represent the uncertainty due to the
inclination angle.  For the three objects whose jet expansion times
are determined from proper motions, $t_j$ is independent of the
inclination and the distance, and the error bars in the figure reflect
the formal errors from the proper motions.  For these objects, the
distance does contribute to the uncertainty in $t_t$, but in all cases
it is well determined (to within $\sim \pm15$\%, see references in
Table~1), and for simplicity this additional contribution is omitted
in the figure.

A third effect, which is important for He~3-1475 and AGFL~618, arises
because their equatorial regions are only marginally resolved by the
observations so the dimensions of the tori are not well determined. In
both cases the values for the torus parameters are taken from models
of the CO emission (see appendix). The errors are difficult to
quantify but are probably within a nominal overall factor of two in
$t_t$, and these are shown as gray error bars in the figure.

The expansion times of the jets and tori shown in Fig.~1 span a large
range, from $\sim 100$~yr to several 1000~yr. In spite of the
heterogeneity of the data and the uncertainties in individual cases,
the figure shows that the ages of the jets and the ages of the tori
are well correlated, and that the data points lie not far from the
line $t_j=t_t$. The implication is that the torus and jets in each
object were ejected nearly simultaneously.

It is also of interest to note that the ejection ages are roughly
ordered according to the degree of evolution indicated by the spectral
type of the star and/or the development of the nebula: i.e., from
youngest to oldest corresponds to AGB stars, through proto-PNe
(typically B-type spectra), to true PNe. This is consistent with all
ejections (except perhaps M2-9 whose status is uncertain, see \S2)
happening in the final stages of the AGB phase, or shortly thereafter.

\subsection{Jet-Lag}

The jets and tori have similar ages, but the data points in Fig.~1 lie
systematically below the line $t_j = t_t$.  The effect appears more
pronounced for the younger systems in the log-log plot, but it is also
present for the older systems as well. The data in a few cases are
consistent with $t_j = t_t$ within the uncertainties, but for all
objects our best estimates of the time difference $\Delta t = t_t-t_j$
(column 9 in Table~1) are positive.  This ordering has previously been
noted in individual cases \citep[e.g.,][]{for98,hug00,san04}, but the
result for the ensemble as a whole makes it much more concrete,
especially in view of the fact that $t_j$ refers to the heads of the
jets, and $t_t$ refers to the mean radius of the tori (or some similar
measure) in most cases. In fact, given the diversity of the data used
to compile the figure, the uniformity of the effect is striking.

There are two likely explanations of the observed effect that need to
be considered, and we discuss each in turn.  The first is that the
values of $t_j$ and $t_t$ are good approximations to the actual travel
times of the jets and tori (as previously assumed), but the jets and
tori are not exactly simultaneous. The observed effect is produced if
the jets are slightly younger than the tori, i.e., there is a small
but systematic delay in the launching of the jets relative to the
tori. If so, this delay, or jet-lag, is given by the value of $\Delta
t$.  It might correspond, for example, to a power-up time for the
jets, or an accretion time scale (\S4.2.2). Inspection of Table~1
(column 9) shows that the largest value of $\Delta t$ is for the
oldest system (KjPn~8), but the others are comparable, within the
large uncertainties.  The median jet-lag is 300~yr.

The second explanation to consider is that the jets and tori are
ejected simultaneously, but the expansion times differ from the actual
travel times. The expansion times of the massive molecular tori are
almost certainly good approximations to the travel times for reasons
given earlier. In the event that they are formed deep within the star,
they could be decelerated by overlying layers, but this is unlikely to
affect the timing over the large distances and long time intervals
accessible to the observations. We therefore confine our attention to
the jets. These move through the circumstellar environment at high
velocity, and from a theoretical point of view their kinematic
behavior is uncertain, so we consider various possibilities in turn.

If the ejection of a jet is short-lived, the jet head may act as a
bullet, in which case it would travel at a constant velocity, or
decelerate to the extent that it interacts with ambient circumstellar
gas.  The deceleration of He~3-1475 has been considered by
\citet{rie03} and is estimated to be relatively small, although the
theory of bullet interactions with ambient gas is not fully
developed. More generally, if the head decelerates, it can be seen by
writing the actual travel time as $t = \int  dr/v(r)$ that the value of $t_j$
given by the current (observed) value of $r_j/v_j$ (or the equivalent proper
motion  $\phi/\dot{\phi}$) overestimates the 
travel time.  Therefore, if the ejections of the jets and tori
were simultaneous, the apparent ages of the jets would be older than
the tori, an effect opposite to that seen in Fig.~1.  We conclude that
deceleration does not dominate the observations, and if it were to
play any quantitative role, the true jet-lag would be larger than that
tabulated in Table~1.

A different situation occurs if the jet is continuous over an extended
interval of time.  To illustrate this case, we consider a jet driven
through the circumstellar gas by a radial, momentum-conserving, fast
outflow, with a narrow, fixed, opening angle. For an envelope with a
density that varies as $r^{-2}$, which corresponds to a constant mass
loss rate, the velocity of the jet head is a constant to first order
\citep{lee01}. However, according to the simulations of \cite{lee03},
the effects of the motion of material along the sides of the jets may
lead to a small acceleration.  The velocity of the jet head may then
be written as $v = u_j(1+\alpha t)$, where $u_j$ is the initial
velocity and $\alpha$ is a parameter with a typical value $\sim
10^{-3}$~yr$^{-1}$ for the simulations they present.  If the jets and
tori were ejected simultaneously, we can identify $t$ with $t_t$ and
would expect the apparent ages of the jets to be smaller than the ages
of the tori by $\Delta t$ given by:
\begin{equation}
\frac{\Delta t}{t} = \frac{\alpha t }{2(1+\alpha t)} .
\end{equation}
This expression has values of $\alpha t/2 \sim 0$, 0.25, and 0.5, for
$\alpha t \ll 1$, $\alpha t = 1$, and $\alpha t \gg 1$, respectively.
For an ensemble of objects, we then expect ${\Delta t}/{t}$ to
range from near zero for young systems, up to 0.5 for very old
systems. The observed values are listed in column (10) of
Table~1. They show the opposite behavior, with the largest values for
the youngest systems. It is not possible to rule out acceleration in
any individual case, but there is no strong evidence for it from the
ensemble of data. In view of this, and the extensive empirical
evidence for approximately Hubble-like jet flows cited earlier, we
conclude that jet acceleration is probably not the primary effect, and
that the jet-lag sequence is the preferred one.

On the basis that the expansion times are reasonable approximations of
the travel times, we can reconstruct the time evolution of the
torus-jet sequence.  The time line for each object is shown in Fig.~2.
The filled symbols denote the values of $t_t$ already discussed, and
represent the average or characteristic expansion times of the
tori. The horizontal lines represent the time interval over which the
tori are ejected. These are based on the radial extents of the tori
(e.g., the half intensity widths or model fits to the radial extent),
and assuming the ejection takes place at the mean expansion velocity
for each object listed in the Table~1. If the tori are ejected with a
significant dispersion in velocity, the time lines overestimate the
width of the corresponding ejection interval. M~1-92 may be an extreme
case with a velocity gradient that suggests it was ejected in a short
time \citep[][see also Appendix A2]{alc07}. The information on the
partially resolved tori (AFGL~618 and He3-1475) is incomplete. The
triangles on each line in Fig.~2 denote the times at which the jets
are ejected, and are subject to the uncertainties in timing discussed
above.  For two objects, M 1-16 and He 3-1475, the timing of later
jets (see appendix) is also shown, to illustrate the overall sequence.

The ejection sequences for all the objects shown in Fig.~2 are
qualitatively similar.  There is a rapid build-up of the torus and
then the jets are launched, and may reoccur. As discussed above, our
best estimates suggest a finite lag time for the jets, with a median
of 300~yr. If this is measured with respect to the onset of the torus
ejection, it is probably somewhat longer. The interval between the
jets in the objects with a sequence of jets is comparable to the
jet-lag time.

\section{Discussion}

\subsection{Jets and Tori}
The ejection of jets and tori are among the most important events in
the transition of stars from the AGB to the PN phase. Our finding that
they occur close together in time provides strong evidence that they
are physically related, either causally, or by some underlying process
or event linking the two.

Our additional finding that jets and tori probably occur in a
particular sequence, underscores their connection.  These results,
together with the time-scale for the torus ejection and the time scale
for the torus-jet sequence, provide basic constraints on formation
scenarios.

Despite their connections, it is important to emphasize that the
actual ejections of jets and tori are different in character. The jets
involve high velocity, axial outflows, while the tori involve low
velocity equatorial outflows. The difference in the velocities is
highlighted by the data listed in Table~1. The median velocity of the
jets is 160~\kms\ and the median velocity of the tori is 10~\kms. It
is noteworthy that the latter is comparable to, but below the typical
wind velocity on the upper AGB: the median envelope velocities for
bright carbon stars, infrared carbon stars, and OH/IR stars, are
11.4~\kms, 17.8~\kms, and 13.6~\kms, respectively
\citep[][]{olo03}. In addition, in the individual cases M~1-16
\citep[][]{hug00} and AFGL~618 \citep[][]{san04}, the tori have lower
expansion velocities than the extended AGB envelopes. Thus the torus
ejection, which may also involve a very high mass-loss rate (\S1), is
likely to be a different process from, or at least a modification of,
the regular AGB mass loss.

\subsection{Formation Scenarios}

The linking of jets and tori, and their related time scales,
considerably extend the range of constraints we have on jet
formation. In the following, we comment on how these considerations
affect our evaluation of specific scenarios that have been
proposed. Table~2 is used to help keep track of the different
possibilities.

\subsubsection{Magnetic Winds from Single Stars}

One scenario for jet formation proposed by Garc{\'{\i}}a-Segura and
co-workers is a collimated wind from single stars in the proto-PN
phase. Their models \citep[e.g.,][]{gar05} do not appear to naturally
give rise to the ejection of a torus close in time with the jets.  The
equatorial density enhancements that are featured in some of their
simulations are input separately, and do not appear to constitute a
discrete ejection. In view of the close connection between jets and
tori presented in this paper, the single star models do not currently
give a good description of the observations.

\subsubsection{Companion Accretion Disks}

A second class of scenarios for jet formation occurs in binary systems
where part of the matter lost in the wind of the evolving AGB star
(the primary) is captured by a binary companion (the secondary) and
forms a disk \citep[][]{mor87,sok00}. The disk in turn forms jets.
The jet launching mechanism is not known for certain, but is probably
similar to that in young stellar objects \citep[e.g.,][]{fra04}.

There is growing evidence for binary interactions in AGB stars with
high mass loss rates \citep[e.g.,][]{mau06}, and the binary accretion
disk scenario is consistent with our findings on several points.  (1)
The presence of a close binary provides a mechanism to enhance the
mass loss of the primary, and to preferentially confine the mass loss
to the orbital plane. (2) If a discrete torus forms, the increase in
mass loss produces an increase in the accretion rate of the disk;
assuming that jets are enhanced or perhaps triggered by the increased
accretion, this model provides the basis of a causal link between tori
and jets.  (3) This causal relationship also has the potential to
account for the delay in launching the jets.

The delay due to the travel time of material between the primary and
the secondary is expected to be small under most circumstances, e.g.,
$\Delta t = 5$~yr for a separation of 10~AU and a gas velocity of
10~\kms. However, a significant delay could arise because it takes a
finite time for matter accreted by a disk to spiral into the inner
regions where the jets are presumably launched.  In the present
scenario, the increased accretion rate of the disk will occur during
the build-up of the torus, so that the jets will be delayed by a time
comparable to the accretion time of the disk. This is given by the
expression $t_{\nu} = R^2/\nu$, where $R$ is the radius of the disk
and $\nu$ is the viscosity. In the $\alpha$-prescription
\citep{sha73}, this can be parameterized as:
\begin{equation} 
t_{\nu} = 160\,\mathrm{yr} \, \left(\frac{\alpha}{0.1}\right)^{-1} 
\left(\frac{R}{1\, \mathrm{AU}}\right)^{3/2} \left(
\frac{M_2}{1\,M_\odot} \right)^{-1/2} \left( \frac{H/R}{0.1}
\right)^{-2}  ,
\end{equation} 
where $\alpha$ is the usual viscosity parameter, $M_2$
is the mass of the companion, and $H$ is the scale height of the disk.

The sizes of the disks are unknown, but simulations of accretion by
relatively close companions \citep[][]{the93,mas99} and tidal
truncation \citep[][]{pas89} suggest disks with radii $\sim 1$~AU may
be formed. For disks of this size and reasonable values for the
companion mass and disk thickness, it can be seen that if $\alpha$ is
as low $10^{-3}$--$10^{-4}$ as found by \citet{bel94} to reproduce the
flaring time of FU Ori stars, the accretion times would be $\ga
10^4$~yr. This kind of time scale is completely ruled out by the short
jet-lags observed. For a typical jet-lag of $\sim 300$~yr, a disk of
radius 1~AU requires $\alpha \ga 0.05$. Thus, if this model proves
relevant, the jet-lag has the potential to probe properties of the
disks that are currently unobservable in any other way.

The points outlined above are consistent with the binary accretion
disk model. In addition, if M~2-9 proves to be a symbiotic system with
a detached companion (see \S2), it adds support to this
interpretation. There is, however, one aspect of the observations that
is not naturally explained by this scenario, and that is the ejection
of a torus. Enhanced mass-loss is certainly a feature expected in
close binary systems, but the formation mechanism of a discrete torus
is unclear.

In their binary model for PNe with narrow waists, \citet{sok00} argue
that the tori are formed by the effects of a collimated fast wind
(i.e., the jets) from the companion, by concentrating the AGB wind
toward the equatorial plane. If the fast wind extends over a wide
angle in latitude and forms a hot bubble, it may contribute to the
equatorial concentration, but the simulations of \cite{gar04} with
more directed jets show mainly local enhancements around the flow
cavities and no clear build up of a large scale torus. In any event
the relatively sudden appearance of low velocity tori with large
masses (and large mass loss rates) cannot be explained by simply
compressing the gas in latitude.  A trigger or event to produce the
tori is much more likely.  One possibility is magnetic expulsion along
the lines discussed in the next section, powered by spin-up of the
envelope, and triggered at a point when the overlying mass of the AGB
envelope is reduced to a critical level. This and other ways to
generate a torus within this scenario warrant further attention.

A variant of the binary accretion disk scenario is one in which a disk
first forms around the companion and blows jets (e.g., as a result of
wind accretion or Roche lobe overflow of the primary), and then enters
the primary in a common envelope and ejects a torus (see next
section). This sequence of jets followed by a torus is ruled out by
the observations.

\subsubsection{Magnetic Effects in Common Envelopes}

A range of scenarios for jet formation occurs where there is direct
interaction between a close binary companion and the AGB star in a
common envelope phase \citep{ibe93}. In this phase, the companion is
engulfed by the AGB star: it may spiral in toward the center on a
rapid time scale $\la 10$~yr, and if the mass of the companion is
above $\sim 0.1$~M$_{\odot}$, it can deposit enough energy to eject a
significant portion of the AGB envelope \citep[e.g.,][]{sok94,nor06}.
Simulations show that the envelope is ejected on a short time scale
($\sim 1$~yr), mainly in the equatorial plane of the system
\citep{san98}. The ejection, therefore, provides a natural mechanism
for the formation of an expanding torus.

During the interaction, the envelope may be spun up by the secondary,
and this can enhance the magnetic field by dynamo action to produce a
jet, with a short lifetime $\la 100$~yr \citep{nor06}. If both spin-up
and ejection occur during the common envelope phase, this would
generate jets and a torus close in time as observed. It remains to be
seen from detailed simulations whether the jet-lag sequence is a
natural feature of this scenario.

A related scenario discussed by \citet{mat06} is one in which
rotational shear between the core and envelope builds up the magnetic
field until it explodes in both the polar and equatorial
directions. This case is attractive because the jets and torus are
likely to form close in time, as observed, and the magnetic field in
the torus may have observable characteristics \citep{hug05,sab07}.
Again, it remains to be seen from detailed simulations whether the
jet-lag sequence is a natural feature of this scenario.

In either the hydrodynamic or the magnetic expulsion of the equatorial
gas, a very important constraint provided by the ensemble of observations
in Table~1 is the low velocity of the tori (Table~1). The ejection of
matter from the deep potential well of a close binary or in a magnetic
explosion likely leads to a range of ejection velocities and this
point needs to be evaluated in realistic simulations. If the
simulations are unable to produce coherent low velocity tori, these
formation mechanisms could be ruled out.

Several other considerations bear on this class of scenarios. In the
case of V~Hya (Table~1), \cite{bar95} have found that the photospheric
lines are broadened, and interpret this as evidence that the star is
currently in the common envelope phase. If this is so, the radially
extended torus \citep{hir04} implies an extended interval of
equatorial mass-loss of $\sim 1500$~yr as shown in Fig.~2. This is
much longer than the ejection times found in simulations of common
envelope ejection noted above. On the other hand, \cite{kna99} find
evidence for a 17~yr period in V~Hya, which may indicate a detached
binary system.

In the case of M~1-92 (Table~1), \cite{alc07} interpret their
observations to indicate that the jets and torus are exactly
simultaneous, and therefore support the magnetic explosion model. For
the best available estimate of the inclination angle of the system
that we adopt (see appendix), the jets are marginally younger than the
torus, as in all the other objects, so there is no compelling reason
that they must be simultaneous. It is also true that the magnetic
expulsion models are at very early stages of exploration; they may not
in fact give rise to exactly simultaneous polar and equatorial
ejections and may be consistent with the jet-lag picture.

\subsubsection{Primary Accretion Disks}

A further class of scenarios that may produce jets involves a binary
companion that forms an accretion disk around the primary during or
after the common envelope phase.

If the companion is of low mass (brown dwarf or planet), it enters the
AGB star and spirals in toward the core, where it is gravitationally
shredded to form a disk that may blow jets through the envelope
\citep{sok96,rey98,nor06}.  A very low mass secondary has insufficient
energy to eject much of the AGB envelope during the spiral-in process,
so it does not naturally lead to the formation of a massive
torus. However, the primary may go on to later eject the envelope,
possibly with an equatorial enhancement. In this scenario, there is no
particular co-ordination of jets and torus, and it yields a sequence
of jets followed by a torus, which is opposite to that observed in all
the objects discussed here. This scenario can be rejected.

In a variant of this scenario \citep{nor06}, the mass of the secondary
may be enough to eject part of the AGB envelope in a torus before the
companion is shredded to form a disk.  This leads to a torus-jet
sequence, as observed.  The time sequence is likely to be short
because the spiral-in and break-up times are rapid, but a viscous
jet-lag of the form given by equation (2) may be relevant, if the
viscosity parameter is low enough to compensate for a small disk
radius.

An alternative scenario with an accretion disk around the primary
occurs when the secondary is a main sequence star of mass $\ga
0.1$~M$_{\odot}$. In this case, the envelope of the AGB star is
ejected in the common envelope phase as a torus; the spiral-in process
of the secondary comes to halt, but it may subsequently undergo Roche
lobe overflow to form a disk around the primary \citep{sok94}. This
produces a clear torus-jet sequence as observed. However,
there are two problems with this scenario. First, the timescale
between the torus ejection and the disk-jet formation is governed by
the thermal response time of the secondary, which is 10$^3$--10$^4$~yr
\citep{sok94}. This is somewhat too long compared with the typical
jet-lag sequence that we find from the observations.  The second
problem is that the jets in this scenario are produced after the
common envelope phase. It seems unlikely that the coolest objects that
we consider (which have already formed jets) are post common-envelope;
this may also apply to all the objects in the sample given the ages of
their jets and their likely evolution times across the H-R
diagram. Except for the ejection plus break-up case noted above, these
primary disk scenarios are not well matched to the observations.

It is interesting to note that \cite{mit06} have recently reported on
the kinematics of the PN Abell~63, which has a close binary central
star and must have gone through a common envelope phase. \cite{mit06}
find that the lobes (jets) in Abell~63 are older than the nebular rim
which forms the inner edge of a cylindrical torus. They conclude that
this is consistent with jets formed after a common envelope (from a
disk around the primary core) but before formation of the main
nebula. As remarked above, the sequence of jets followed by a torus is
the opposite of all the cases discussed in this paper.  It could be
that Abell~63 is different. On the other hand, the jets are very old
(13,000~yr), so it is unclear if simple kinematics can be used to
infer such long travel times. In addition, the age of the rim is
determined from observations of ionized gas whose dynamical evolution
during the formation of the nebula is uncertain.  In view of these
caveats, the conclusions of \cite{mit06} on the jet sequence in
Abell~63 need to be viewed with some caution.

\subsubsection{Summary Evaluation}

From the above discussion, it can be seen that there are quite a
number of scenarios that might potentially produce jets in proto-PNe,
and it may be that more than one of them actually occurs. However, the
properties of the objects discussed in this paper, with well defined
jets and molecular tori, are sufficiently uniform that a common
mechanism seems most reasonable.

On this basis, our discussion shows that the timing observations place
quite stringent constraints on the physical picture, as summarized in
Table~2. Some specific scenarios for jet formation, in which the jets
and tori occur in the wrong sequence, or are unconnected, or have time
scales that are too long, are ruled out or made implausible. The
relatively sudden ejection of a discrete torus followed by jets on a
short time scale most naturally fits the types of scenario in which a
companion directly interacts with the central star. The physical
situations in these scenarios are quite different, but theoretical
exploration has not yet gone much beyond order of magnitude
estimates. More realistic simulations are needed to predict the
detailed characteristics of specific models so that we can
discriminate among the possibilities.

\section{Conclusions}

The objects described in this paper cover a range in evolution from
AGB stars to young PNe, but they all exhibit high velocity polar jets
and dense equatorial molecular tori. From analysis of the timing of
the jets and tori we reach the following conclusions.

First, the launching of the jets is approximately simultaneous with
the ejection of the tori. This implies a close connection between the
two, and provides important information on the physical system in
which the jets are formed.

Second, there is evidence for a torus-jet sequence. In some cases the
observations are consistent with close or exact simultaneity, but the
ensemble of data suggest a torus-jet sequence with a median jet-lag
$\sim 300$~yr. The main uncertainty here is the possible role of jet
acceleration. This has a somewhat similar effect as a delay, but is
probably not the dominant effect.

Third, the ejection velocities and time scales provide further
constraints on the physical picture. The low expansion velocities of
the tori in particular may pose a challenge for models in which tori
are ejected from the deep potential well of a common envelope.

These findings already rule out or make implausible several jet
formation scenarios.  Overall, the relatively sudden ejection of a
discrete torus followed by jets on a short time scale most naturally
fits the types of scenario in which a companion interacts with the
central star.  More detailed simulations of specific models are needed
to help us discriminate among the possibilities.

\acknowledgements
It is a pleasure to acknowledge collaborative programs with
R.~Bachiller, P.~Cox, and T.~Forveille, the results of which form the basis
of this paper.  I thank E.~Blackman, T.~Forveille, and
A.~Frank for comments on the manuscript. This work has been supported
in part by NSF grant AST 03-07277.

\appendix

\section{Details of the Observations}

This appendix provides further information on the origin and
determination of the quantities given in Table~1, and shown in Figs.~1
and 2.

\subsection{Inclination Angles}

The inclinations of the jets and tori to the line of sight ($\theta_j$
and $\theta_t$) are important for estimating the expansion
time scales because of projection effects. For the jets, the observed
radial extents and velocities are $r_j \sin \theta_j$ and
$v_j \cos \theta_j $, respectively, where $r_j$ and $v_j$ are the
actual (un-projected) values. For the tori, the radii $r_t$ are seen
directly on the sky at the systemic velocity, and the maximum observed
radial velocities are $v_t \sin \theta_t$.  The expansion time scales,
$t_j$ and $t_t$, are therefore sensitive to the inclination angle, and
vary as $\cot \theta_j$ and $\sin \theta_t$, respectively.  Exceptions
for which the expansion time scales are determined in a different way
are noted below.

The adopted inclination angles for each source are given in Table~3,
together with estimates of the uncertainty. In five sources there are
independent estimates of $\theta_t$ and $\theta_j$, and in the others
there are estimates of only one: for these we assume that $\theta_t =
\theta_j$. This is reasonable on the basis of the similarity where
there are independent estimates, and by the fact that many (though not
all) images show that jets are roughly aligned with the axes of the
tori when seen in projection.

The uncertainties adopted for the inclination angles given in Table~2
are of two types. In specific cases noted in the footnotes to the
table, the values of $\Delta\theta$ are taken from the references
cited. For the others we have made rough estimates, based on the the
observations and, where relevant, on the agreement between different
estimates: for simplicity in these cases we have adopted values for
$\Delta\theta$ of $\pm5\arcdeg$ or $\pm10\arcdeg$. For the sources
where we assume $\theta_t = \theta_j$, the errors in Fig.~1 are
correlated; and for the three objects in which $t_j$ is estimated
directly from proper motions, the results are independent of the
inclination.

\subsection{Notes on Individual Sources}

KjPn~8.-- $\theta_t$ is re-determined from the velocity strip
maps of the torus \citep{for98}, and agrees with $\theta_j$ from the
jet kinematics given by \citet{mea97}. $t_j$ is from optical proper
motions of the jets, $34\pm3$~mas~yr$^{-1}$ \citep{mea97}.

M~1-16.-- The adopted value of $\theta_j$ \citep[from][]{sch92} is at
the top of the range given by \cite{cor93}; this value is preferred
because the jets align with the torus axis on the sky and this is the
closest value to $\theta_t$, which is well determined.  The inner edge
of molecular torus is not well resolved by the CO observations; the
limit in Fig.~2 corresponds to the radius of the small ionized nebula.
$t_j$ for the two additional jet components in Fig.~2 are 1050~yr and
740~yr \citep{sch92}.

M~2-9.-- $t_j$ is from optical proper motions of the jets,
$51\pm7$~mas~yr$^{-1}$ \citep{sch97}.

M~1-92.-- The adopted value of $\theta_j=57\pm5\arcdeg$ \citep{sol94}
is from a geometrical method using Doppler shifts in the jets.  $t_t$
and $t_j$ are based on the equatorial and polar velocity gradients (12
and 7.6~\kms~arcsec$^{-1}$, respectively) given by \cite{alc07}; using
the gradient for the torus gives a value of $t_t$ that depends on $\tan
\theta_t$. The gradient suggests a torus ejection event, although the
jet outflows in this source are particularly wide-angle and extend to
low latitudes so that the torus may be wind-swept; this may
account for the gradient. \cite{alc07} assume the torus and jets are
ejected at the same time, which requires an inclination angle of
$51.5\arcdeg$ to get the same expansion time scales. In Figs.~1 and 2
we use the independent value of \cite{sol94} given above.

M~2-56.-- The inner rim of torus not well resolved; the limit in
Fig.~2 corresponds to half the beamsize. 

He~3-1475.-- The data for the torus are uncertain because it is
marginally resolved by the CO observations. The adopted values are
from the CO model by \cite{hug04}. Other rough estimates give
comparable values for $t_t$. For example, the strong, low velocity OH
masers are spread over an area of radius $\sim {0}\farcs{5}$ and a
velocity of $\sim 18$~\kms\ \citep{zil01}; this gives $t_t \sim
1000$~yr.  $t_j$ is from optical proper motions of the jets,
$12.6\pm1.1$~mas~yr$^{-1}$ \citep{rie03}.  The younger jets shown in
Fig.~2 have $t_j = 446$~yr, from proper motions for knots NW2 and SW2
given by \cite{rie03}.
 
V~Hya.-- $r_t$ is the half power radius at the systemic velocity in
the CO 3--2 line from the maps by \cite{hir04}. The inner rim is not
well resolved, the limit in Fig.~2 corresponds to half the beamsize.

AFGL~618.-- The data for the torus are uncertain because it is
marginally resolved by the CO observations. The adopted values are
from the CO model by \cite{san04} with velocity $\propto$ radius; the
inner rim is not resolved. Roughly the same value of
$t_t$ ($\sim 400$~yr) is obtained independently from an expanding
torus model of the dense inner regions ($r_t \sim {0}\farcs{75}$, $v_t
\sim 5$--12~\kms) observed in HC$_3$N \citep{par04}.  $\theta_j$ is
for the single jet $b$ in the nomenclature of \cite{tra02} and $v_j$
is the H$_2$ velocity for same jet \citep{cox03}. 

$\pi^1$~Gru.-- Data for the torus are from the CO model by
\cite{chi06}; $r_t$ is the geometric mean of the inner and outer model radii.

\clearpage

\begin{deluxetable}{lcccccccccc}
\tablecaption{Properties of the Jets and Tori}
\tabletypesize{\scriptsize}
\tablehead{
 \colhead{Name} &  \colhead{$d$} & \colhead{$r_t$ } & \colhead{$v_t$ } &
  \colhead{$t_t$ }& \colhead{$r_j$ } & \colhead{$v_j$ }&
  \colhead{$t_j$ }&  \colhead{$\Delta t$} & $\Delta t/t_t$ &\colhead{ref} \\
   & \colhead{(kpc)} & \colhead{(10$^{16}$ cm)} &
  \colhead{(km\,s$^{-1}$)} & \colhead{(yr)} & \colhead{(10$^{17}$ cm)} &
  \colhead{(km\,s$^{-1}$) }& \colhead{(yr)} &  \colhead{(yr)} & &
}
\startdata
KjPn~8   &  1.60 & 9.3 & 5.9   & 5040 & 34.6 & 314 & 
 3380\tablenotemark{a} & 1660 & 0.33 & 1, 2         \\
M~1-16   &  1.80 & 6.2 & 9.8  & 2000 & 17.8 & 350 &
 1610\tablenotemark{b} &  390  & 0.20 & 3, 4\\
M~2-9    &  0.64 & 2.9 & 7.0 & 1300 & 5.93  & 164 & 
 1170\tablenotemark{a} & 130 & 0.10 & 5, 6 \\
M~1-92   &  2.50 & 2.6 & 5.5 & 1520 & 2.21  & 69  &
 1010\tablenotemark{c} & 510 & 0.34 &  7, 8   \\
M~2-56   &  2.10 & 4.4 & 8.0 & 1750 & 4.27  & 128 &
 1060\tablenotemark{c} & 690 & 0.39 &  9  \\
He~3-1475&  5.80 & 3.9 & 14  & 878  & 10.4 & 527 &
 611\tablenotemark{a} & 267 & 0.30 & 10, 11      \\ 
V~Hya    &  0.38 & 2.7 & 16  & 543  & 1.29  & 161 &
 254\tablenotemark{c} & 289 & 0.53 & 12   \\
AFGL~618 &  0.90 & 1.6 & 12  & 422  & 0.97 & 222 &
 139\tablenotemark{c} & 283 & 0.67 & 13, 14, 15   \\
$\pi^1$~Gru& 0.15& 1.2 & 11  & 354  & 0.09  & 55  & 
 54\tablenotemark{c} & 300 & 0.85 & 16   \\ 
\enddata
 
\tablenotetext{a}{based on optical proper motions}
\tablenotetext{b}{based on optical kinematics}
\tablenotetext{c}{based on molecular line kinematics}
\tablerefs{
(1) \citealt{for98}; (2) \citealt{mea97}; (3) \citealt{hug00}; 
(4) \citealt{sch92}; (5) \citealt{zwe97}; (6) \citealt{sch97}; 
(7) \citealt{buj98}; (8) \citealt{alc07}; (9) \citealt{cas02}; 
(10) \citealt{hug04}; (11) \citealt{rie03}; (12) \citealt{hir04}; 
(13) \citealt{cox03}; (14) \citealt{tra02}; (15) \citealt{san04}; 
(16) \citealt{chi06} 
} 
\end{deluxetable}

\clearpage

\begin{deluxetable}{lcccc}
\tablecaption{Jet-Torus Scenarios}
\tablewidth{0pt}
\tabletypesize{\small}
\tablehead{
 \colhead{Scenario} & \colhead{Rating} & \colhead{Comments} 
}
\startdata
Magnetic wind from single star            & $-$  & jets \emph{and} torus?    \\
Primary mass loss + companion accretion disk  & $\bigstar$ &  discrete torus
ejection?   \\ 
Companion accretion disk + CE ejection & $-$ &  wrong sequence  \\ 
CE ejection + magnetic polar wind    & $\bigstar$  &  jet-lag?   \\
(CE) magnetic polar \& equatorial explosion   &  $\bigstar$  &  jet-lag?   \\
(CE) primary accretion disk + late nebula ejection  & $-$ &  wrong
sequence   \\ 
CE partial ejection + primary accretion disk  &  $\bigstar$   &  jet-lag?   \\
CE ejection + post-CE primary accretion disk (RLOF)   & $-$  &
time scale too long?    \\
\enddata
 
\tablecomments{CE = Common Envelope, RLOF = Roche lobe overflow}

 \end{deluxetable}

\clearpage

\begin{deluxetable}{lcccc}
\tablecaption{Inclination Angles}
\tablewidth{0pt}
\tabletypesize{\small}
\tablehead{
 \colhead{Name} & \colhead{$\theta_t$} & \colhead{$\theta_j$} &
 \colhead{$\pm \Delta\theta$} & \colhead{ref} \\
  &  \colhead{($\arcdeg$)} & \colhead{($\arcdeg$)} & \colhead{($\arcdeg$)} & 
}
\startdata
KjPn~8     & 52  & 53                 & 5                     & 1, 2  \\
M~1-16     & 54  & 45                 & 10                    & 3, 4  \\
M~2-9      & 73  & 75                 & 5                     & 5, 6  \\
M~1-92     & 57\tablenotemark{a} & 57 & 5\tablenotemark{b}    & 7     \\
M~2-56     & 73  & 73\tablenotemark{a}& 2\tablenotemark{b}    & 8     \\
He~3-1475  & 40  & 40                 & 5                     & 9, 10 \\ 
V~Hya      & 30  & 30\tablenotemark{a}& 10                    & 11    \\
AFGL~618   & 58  & 51                 & 10, 4\tablenotemark{b}& 12, 13 \\
$\pi^1$~Gru& 35  & 35\tablenotemark{a}& 10                    & 14    \\ 
\enddata
 
\tablenotetext{a}{Assuming $\theta_j=\theta_t$ }
\tablenotetext{b}{Estimate from reference, others from
  this paper.} 
\tablerefs{
(1) \citealt{for98}; (2) \citealt{mea97}; (3) \citealt{hug00}; 
(4) \citealt{sch92}; (5) \citealt{zwe97}; (6) \citealt{sch97}; 
(7) \citealt{sol94}; (8) \citealt{cas02}; (9) \citealt{bor01}; 
(10) \citealt{hug04}; (11) \citealt{hir04}; (12) \citealt{san04}; 
(13) \citealt{tra02}; (14) \citealt{chi06} 
} 
  \end{deluxetable}


\clearpage


\clearpage





\begin{figure*}
\plotone{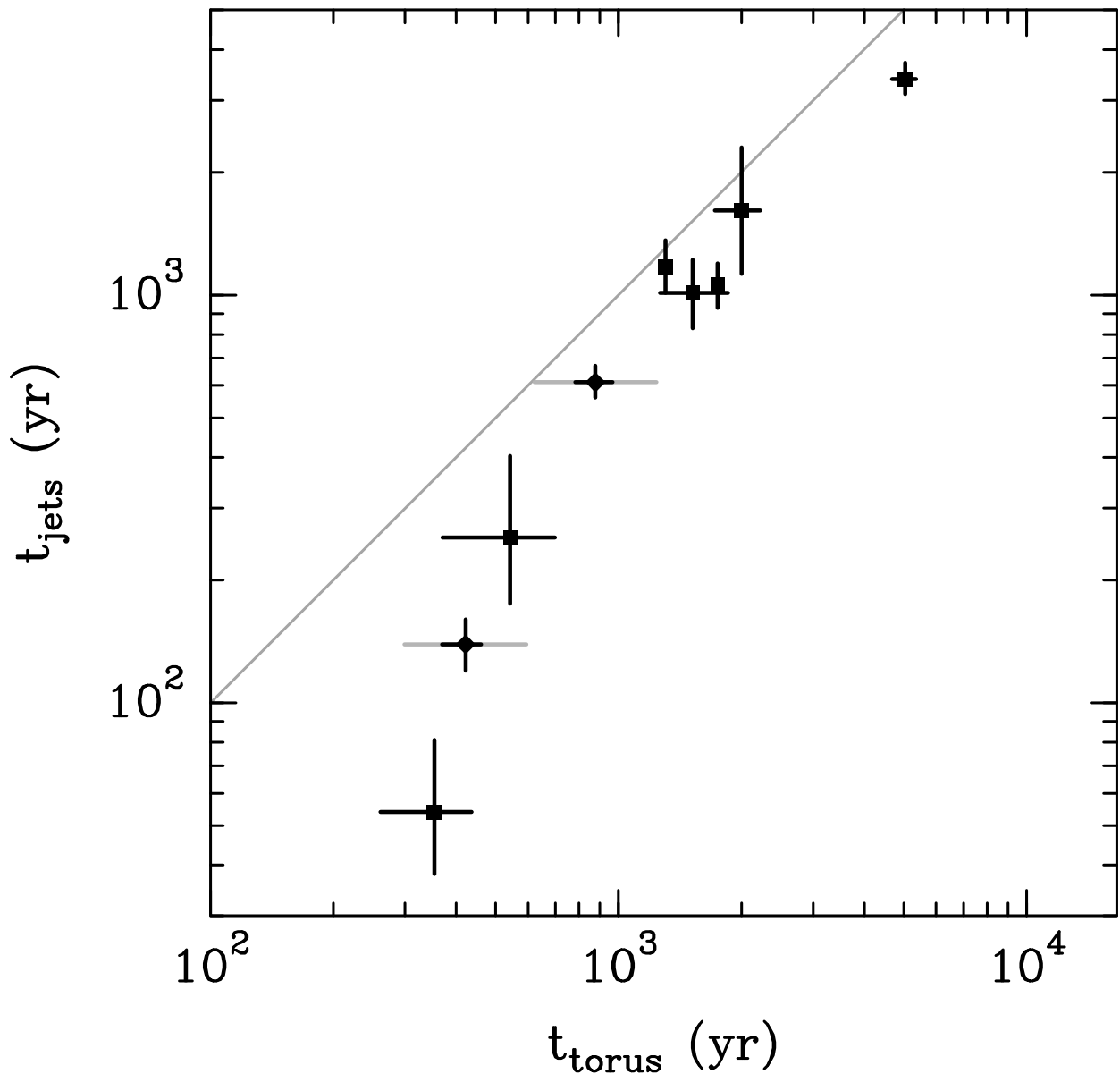}

\figcaption[]{Comparison of the expansion times of the jets and tori.
 The squares denote cases where the tori are resolved, and the
 diamonds where they are marginally resolved. The error bars reflect
 the uncertainties in the inclination angles or proper motions. The
 extra, gray error bars reflect a nominal factor of 2 uncertainty in
 the dimensions of the marginally resolved tori.  The continuous line
 shows the locus on which the expansion time scales are equal.
 }
\end{figure*}

\clearpage 
\begin{figure*}
\plotone{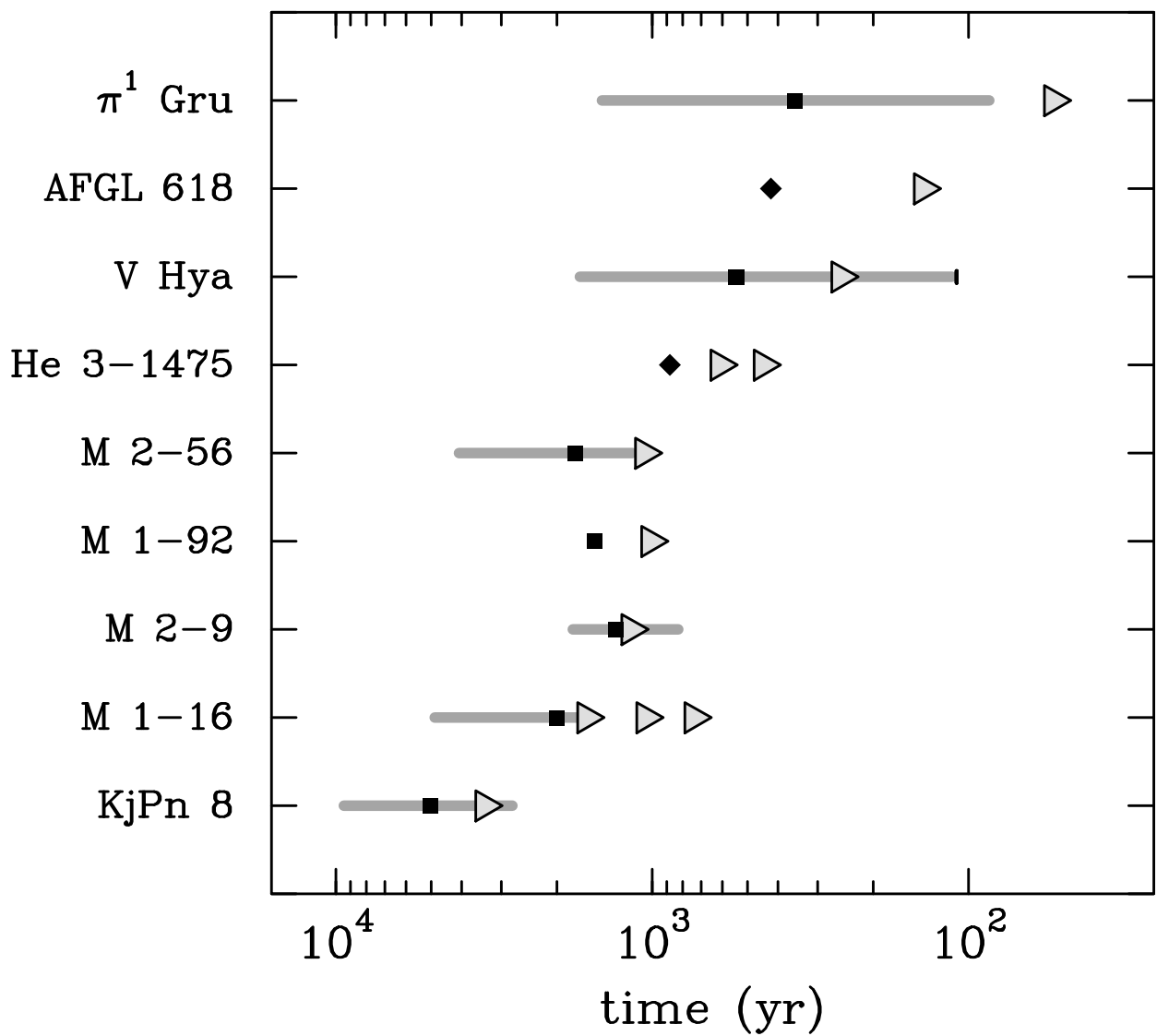}

\figcaption[]{Time sequence for the ejection of jets and tori.  The
  horizontal axis is time in the past, measured from the current
  epoch. The filled squares and diamonds show the fiducial times of
  torus ejection as in Fig.~1. The horizontal lines show the
  characteristic duration of the torus ejection, assuming a constant
  ejection velocity.  In cases where the inner rim of the torus is not
  seen, the time line is truncated with a small vertical line
  corresponding to the resolution limit. Data for the time lines of
  the marginally resolved cases (diamonds) are incomplete. M~1-92 may
  be ejected in a single event. The triangles denote the times of jet
  ejection. Data on the multiple jets in He~3-1475 and M~1-16 are
  given in the appendix.
}
\end{figure*}

\end{document}